\documentclass[amsmath,amssymb,twocolumn,showpacs]{revtex4}
\usepackage{graphicx}
\usepackage{bm}

\newcommand{\pp}[1]{\phantom{#1}}
\newcommand{\be}{\begin{eqnarray}}
\newcommand{\ee}{\end{eqnarray}}

\newcommand{\ba}{\begin{array}}
\newcommand{\ea}{\end{array}}

\begin{document}
\title{
Tensor-product vs. geometric-product coding
}

\author{Diederik Aerts $^1$ and Marek Czachor $^{2,3}$}
\affiliation{
$^1$ Centrum Leo Apostel (CLEA) and Foundations of the Exact Sciences (FUND)\\
Vrije Universiteit Brussel, 1050 Brussels, Belgium\\
$^2$ Katedra Fizyki Teoretycznej i Informatyki Kwantowej\\
Politechnika Gda\'nska, 80-952 Gda\'nsk, Poland\\
$^3$ ESAT-SCD, Katholieke Universiteit Leuven, 3001 Leuven, Belgium
}

\begin{abstract}Quantum computation is based on tensor products and entangled states. We discuss an alternative to the quantum framework where tensor products are replaced by geometric products and entangled states by multivectors. The resulting theory is analogous to quantum computation but does not involve quantum mechanics. We discuss in detail similarities and differences between the two approaches and illustrate the formulas by explicit geometric objects where multivector versions of the Bell-basis, GHZ, and Hadamard states are visualized by means of colored oriented polylines.
\end{abstract}
\maketitle

\section{Introduction}

Coding based on tensor products is well known from quantum information theory and quantum computation. A bit is here represented by a qubit, and a sequence of bits is represented by tensor products of qubits. Nonfactorizable linear superpositions of simple tensor products are called entangled states. The structures of quantum computation are highly counterintuitive and, with very few exceptions, resist common-sense interpretations.

Coding based directly on geometric algebra (GA) is a new concept \cite{AC07} and is rooted in the fact that the set of {\it blades\/} (geometric  Grassmann-Clifford products of basis vectors of a real $n$-dimensional Euclidean or pseudo-Euclidean space) contains $2^n$ elements. Each blade can be indexed by a sequence of $n$ bits and thus is a representation of a $n$-bit number. Linear combinations of blades are called multivectors.
Blades and multivectors possess numerous geometric interpretations and can be visualized in several different ways.

The fact that multivectors can play a role analogous to entangled states and allow for a GA version of quantum computation does not seem to be widely known. Apparently, the first example of a GA version of a quantum (Deutsch-Jozsa \cite{DJ}) algorithm was given in \cite{AC07}. Each step of the algorithm was interpretable in geometric terms and allowed for cartoon visualization (hence the name `cartoon computation'). The construction from \cite{AC07} was quickly generalized to the Simon problem \cite{Simon} in \cite{MO}. The next step, done in \cite{C07}, was a GA construction of all the elementary one-, two-, and three-bit quantum gates. Therefore, the essential formal ingredients needed for a GA reformulation of {\it all\/} of quantum computation are basically ready.

There were some problems with visualizing situations involving more than three bits due to our habit of thinking in three-dimensional terms. Therefore, one of the first motivations for writing the present paper was to introduce a new method of visualization working for arbitrary numbers of bits. Multivectors are here represented by sets of oriented colored polylines. We geometrically interpret distributivity of addition over geometric multiplication and multiplication of a blade by a number. We also explain how to deal with the complex structure (needed for elementary gates and phase factors) without any need of complex numbers. Our representation of the `imaginary number' $i$ differs from the standard representations used in GA. The reason is that the usual representation, where $i$ is an appropriate blade, does not allow to map entangled states whose coefficients are complex into unique multivectors. Our formalism is free from this difficulty.

We begin, in Section II, with a detailed explanation of the GA way of coding, and illustrate each of the concepts by an appropriate geometric object.
In Section III we compare tensor-product and geometric-product coding. We stress important similarities and differences, and outline certain constructions (eg. scalar product and mixed states) that may prove useful in some further generalizations, but at the present stage are just a curiosity. We do not explicitly introduce elementary gates and algorithms, since these can be found elsewhere. Instead, we concentrate on those elements of the GA formalism where some important differences with respect to quantum computation occur (eg. the probabilistic nature of quantum superposition principle vs. deterministic interpretation of superpositions of blades).  Finally, in Section IV, we discuss geometric interpretation of multivector analogues of some important entangled states occurring in quantum information theory.

\section{Geometric-product coding}

The procedure is, in fact, extremely simple and natural. Indeed, consider an $n$-dimensional real Euclidean space, and denote its orthonormal basis vectors by $b_k$, $1\leq k\leq n$. A {\it blade\/} is defined by $b_{k_1\dots k_j}=b_{k_1}\dots b_{k_j}$, where $k_1<k_2<\dots <k_j$. The basis vectors (one-blades) satisfy the Clifford algebra
\be
b_kb_l+b_lb_k=2\delta_{kl}.\nonumber
\ee
The binary number associated with $b_{k_1\dots k_j}$ can be read out by the following recipe: Take a basis vector $b_k$ and check if it occurs in $b_{k_1\dots k_j}$. If it does --- the $k$th bit is $A_k=1$, otherwise $A_k=0$. Check in this way all the basis vectors.

For notational reasons it is useful to denote the blades in binary parametrization by a character different from $b$, say, $c$. The blades parametrized in a binary way will be termed the {\it combs\/} \cite{C07}. Blades and combs are related by the rule
\be
c_{A_1\dots A_n}=b_1^{A_1}\dots b_n^{A_n}
\ee
where it is understood that $b_k^0={\bf 1}$. Combs and blades are, by definition, normalized if they are constructed by taking geometric products of mutually orthonormal vectors.

Sometimes it is useful to be able to speak of {\it real\/} and {\it imaginary\/} blades and combs. This can be achieved by an additional bit, represented by a vector $b_0$. If $b_0$ occurs in $b_{k_1\dots k_j}$, the blade is imaginary --- otherwise it is real.
Denoting logical negation by the prime, $0'=1$, $1'=0$, we define the `imaginary unit' by the complex structure map
\be
i c_{A_0A_1\dots A_n}=(-1)^{A_0}c_{A_0'A_1\dots A_n}.\label{i}
\ee
A general `complex' element of GA can be written in a form analogous to the usual representation of complex numbers
\be
z_{A_1\dots A_n}&=&x_{0_0A_1\dots A_n}+iy_{0_0A_1\dots A_n}\\
&=&x_{0_0A_1\dots A_n}+y_{1_0A_1\dots A_n}.\label{complex c}
\ee
Here $x_{0_0A_1\dots A_n}=x\, c_{0_0A_1\dots A_n}$ and $y_{0_0A_1\dots A_n}=y\, c_{0_0A_1\dots A_n}$ denote elements of GA that are proportional to a real comb $c_{0_0A_1\dots A_n}$, and the proportionality factors $x$, $y$ are also real.

There exists a `mechanical' justification of the `comb' terminology. In order to understand it, let us recall the formula for multiplication of two normalized combs \cite{AC07}
\be
c_{A_0\dots A_n}c_{B_0\dots B_n} &=&
(-1)^{\sum_{k<l}B_kA_l}c_{(A_0\dots A_n)\oplus(B_0\dots B_n)}.
\label{GAr}
\ee
Here $(A_0\dots A_n)\oplus(B_0\dots B_n)$ means pointwise addition mod 2, i.e. the $(n+1)$-dimensional XOR.
Formula (\ref{GAr}) means that the geometric product may be regarded as a projective (i.e. up to a sign) representation of XOR. Figure 1 shows how to mechanically generate a system that behaves according to (\ref{GAr}). The calculation is $c_{10011}c_{01011}=-c_{11000}$.
\begin{figure}
\includegraphics[width=6 cm]{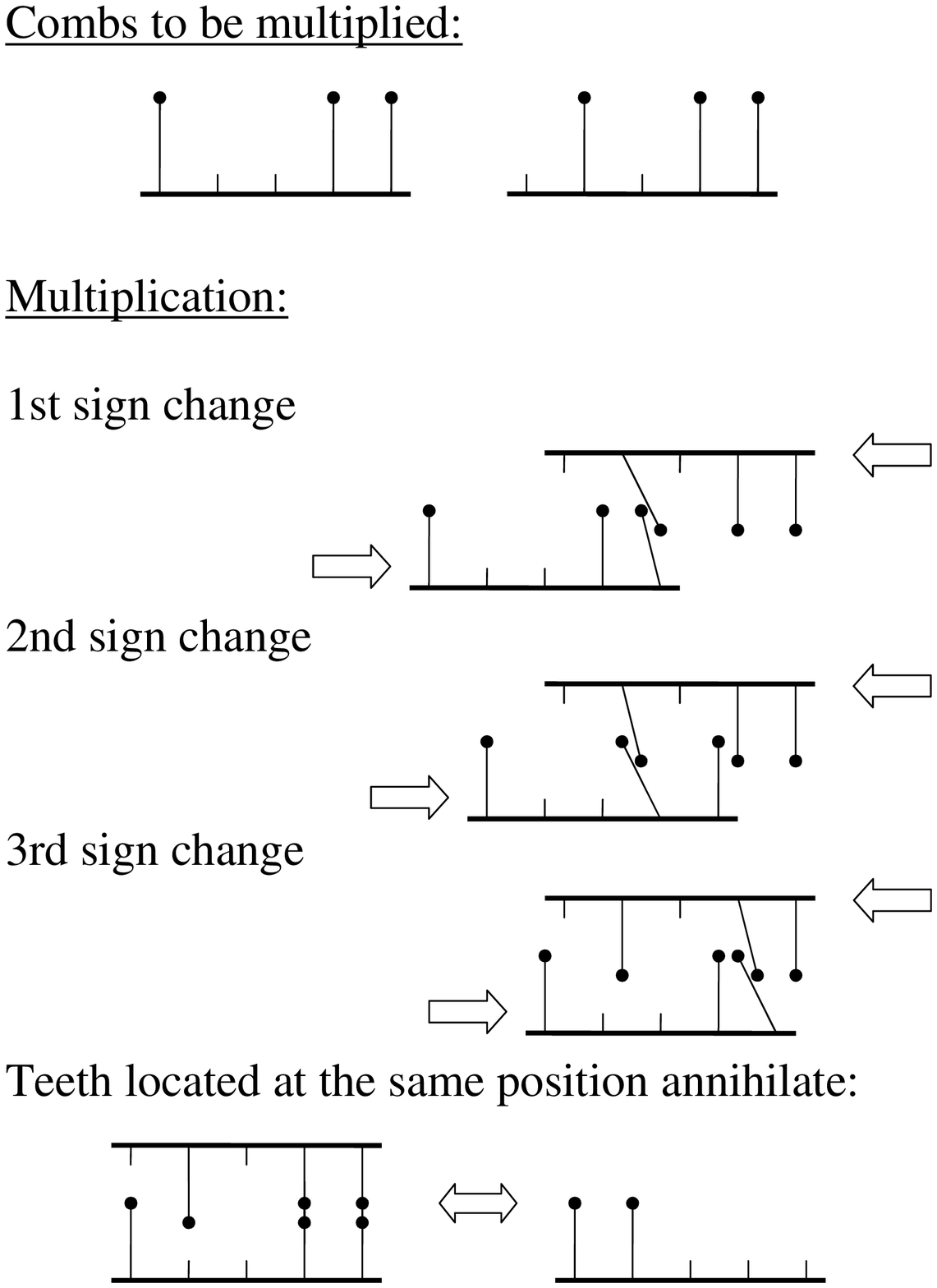}
\caption{Mechanical interpretation of the comb multiplication. (a) Take two combs and flip one of them. Move one comb in the direction of the other. Each time the teeth located in different places meet --- the comb changes its sign. (b) Teeth located at the same position annihilate each another.}
\end{figure}

Combs (and blades) can be visualized in various ways. Geometrically, 0-blades are oriented (`charged') points, 1-blades oriented line segments, 2-blades oriented plane segments, 3-blades oriented volume segments... More precisely, each blade corresponds to an equivalence class of objects. To understand why it is so, consider the algebra of a plane with basis vectors $b_1$, $b_2$. The oriented plane segment
$b_{12}=b_1b_2$ is unaffected by rotations $b_1'=b_1\cos\alpha   -b_2\sin\alpha$, $b_2'=b_1\sin\alpha   +b_2\cos\alpha   $, or rescalings $b'_1=\lambda b_1$, $b'_2=\lambda^{-1} b_2$. Fig.~2 shows an alternative way of visualizing blades in a 6-dimensional Euclidean space (i.e. 6-bit combs), whose dimension is sufficiently counterintuitive. The idea is adapted from a configuration space of three 2-dimensional particles. We distinguish particles by color and tilting of the basis vectors. Oriented segments are represented by oriented multi-color polylines.
\begin{figure}
\includegraphics[width=6 cm]{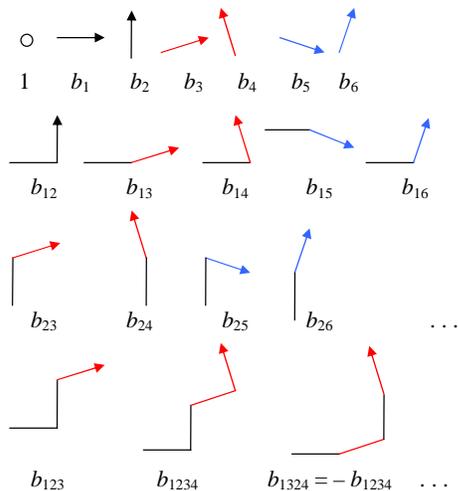}
\caption{(Color online) Colored polyline interpretation of blades.}
\end{figure}
\begin{figure}
\includegraphics[width=8 cm]{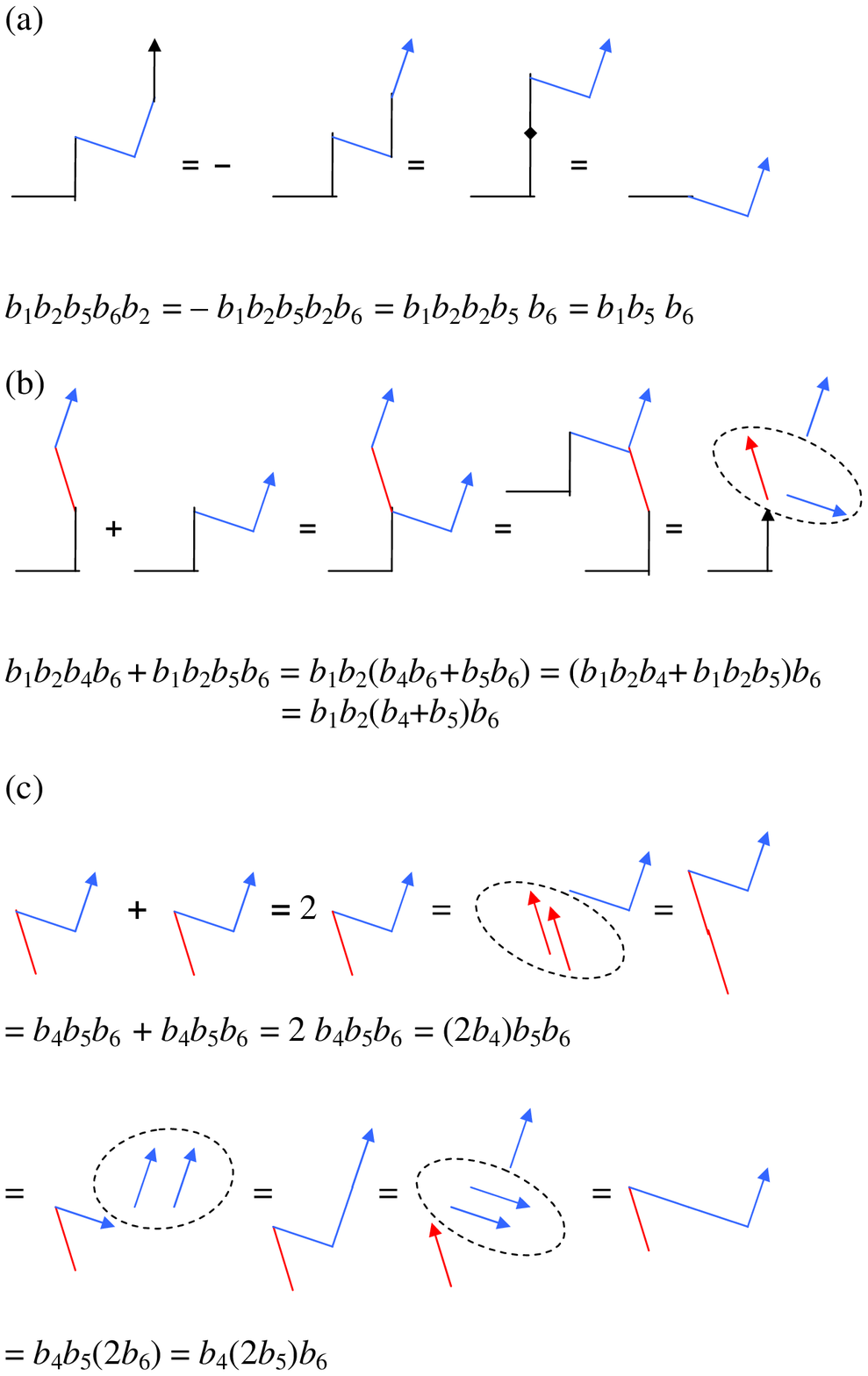}
\caption{(Color online) Arithmetic of blades in 6-dimensional space. (a) Two subsequent segments can be interchanged but orientation (i.e. overall sign) is then changed. Two identical (here unit) segments annihilate when placed one after another. (b) Distributivity of addition of two blades. (c) Multiplication of a blade by a number is derived from distributivity. The blade $2 b_4 b_5 b_6$ is illustrated by means of three different polylines belonging to the same equivalence class.}
\end{figure}

Blades can be added and multiplied by numbers. Fig.~3 illustrates in what sense one can speak of distributivity of addition over geometric multiplication. Addition of two identical blades results in a blade {\it one\/} of whose segments is twice bigger. The three polylines shown in Fig.~3c represent the same equivalence class. Fig.~4 shows a representative of a multivector. Multivectors are `bags of shapes' that differ from visualization to visualization. The concrete multivector from Fig.~4 is $5+1.5 b_1-b_2 +b_1b_4+3b_5b_6+2b_1b_4b_5$. Alternatively, in our binary parametrization, Fig.~4 represents the following superposition of combs
\be
&{}&5c_{000000}+1.5c_{100000}-c_{010000}+c_{100100}\nonumber\\
&{}&\pp=+3c_{000011}+2c_{100110}.
\ee
\begin{figure}
\includegraphics[width=6 cm]{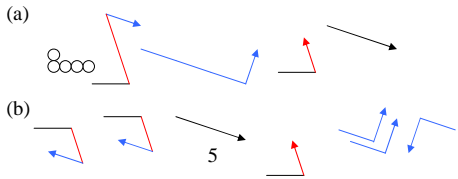}
\caption{(Color online) Two equivalent bag-of-shapes representations of the multivector $5+1.5 b_1-b_2 +b_1b_4+3b_5b_6+2b_1b_4b_5$. The black vector represents $1.5 b_1-b_2$. Representation of scalars by numbers, as in (b), is perhaps more convenient than in terms of `charged points' --- represented by five circles in (a). Yet another representation involves three-dimensional visualization where the scalar part, here 5, is the height of suspension of the plane containing the collection of polylines.}
\end{figure}
As one can see, the oriented colored polyline visualization of multivectors works fine for arbitrary numbers of bits.
This should be contrasted with, say, the representation chosen in \cite{AC07}, where dimensions higher than three led to obvious difficulties.

\section{Similarities and differences between tensor and geometric codings}

Let us now list the basic similarities and differences between coding based on tensor and geometric products.

\subsection{Partial separability}

Two $(k+l)$-bit kets that share the same part of bits, say, $|A_1\dots A_kB_1\dots B_l\rangle$ and
 $|A_1\dots A_kC_1\dots C_l\rangle$, possess the following partial separbility property
 \begin{widetext}
 \be
 \alpha |A_1\dots A_kB_1\dots B_l\rangle+\beta|A_1\dots A_kC_1\dots C_l\rangle
=
|A_1\dots A_k\rangle\big(\alpha |B_1\dots B_l\rangle+\beta|C_1\dots C_l\rangle\big).
  \ee
The property is essential for teleportation protocols.
In geometric-product coding we have an analogous rule,
\be
\alpha c_{A_1\dots A_kB_1\dots B_l}+\beta c_{A_1\dots A_kC_1\dots C_l}
=
  c_{A_1\dots A_k0_1\dots 0_l}\big(\alpha c_{0_1\dots 0_k B_1\dots B_l}+\beta c_{0_1\dots 0_k C_1\dots C_l}\big),
\ee
and thus teleportation protocols can be formulated in purely geometric ways.

Since the link between blades and combs can be written as
$
c_{A_1\dots A_n}=b_1^{A_1}\dots b_n^{A_n}
$
the above rule means simply that
\be
\alpha b_1^{A_1}\dots b_k^{A_k}b_{k+1}^{B_1}\dots b_{k+l}^{B_l}
+
\beta b_1^{A_1}\dots b_k^{A_k}b_{k+1}^{C_1}\dots b_{k+l}^{C_l}
=
b_1^{A_1}\dots b_k^{A_k}
\big(
\alpha
b_{k+1}^{B_1}\dots b_{k+l}^{B_l}
+
\beta
b_{k+1}^{C_1}\dots b_{k+l}^{C_l}
\big).
  \ee
Hence, yet another notation is possible
\be
\alpha c_{A_1\dots A_kB_{k+1}\dots B_{k+l}}+\beta c_{A_1\dots A_kC_{k+1}\dots C_{k+l}}
=
c_{A_1\dots A_k}\big(\alpha c_{B_{k+1}\dots B_{k+l}}+\beta c_{C_{k+1}\dots C_{k+l}}\big).
\ee
\end{widetext}
Here we are making use of the fact that the number of zeros occuring in combs such as $c_{A_1\dots A_k0_1\dots 0_l}$ is a matter of convention: It reflects the freedom of looking at an $n$-dimensional Euclidean space from the perspective of higher dimensions, and treating it as an $n$-dimensional subspace of something bigger.

The notions of product and entangled states can be introduced in the GA formalism in exact analogy to the quantum case.

\subsection{Phase factors}

Complex phase factors play a crucial role in quantum computation and are responsible for interference effects.
An analogous structure occurs also in our geometric formalism, but we first have to comment on the meaning of $i$.

In geometric algebra it is usual to treat $i$ as a bivector. Indeed, GA of a plane consists of $1$, $b_1$, $b_2$, and $b_1b_2$. The latter satisfies $(b_1b_2)^2=-1$, and thus `complex numbers' are often represented in a GA context by multivectors of the form
$x\,1+y\,b_1b_2$, where $x$, $y$ are real. This type of complex structure is employed in \cite{Havel-Doran-Furuta}.

However, there is a simple reason why such a type of `$i$' is not applicable in our formalism.
For assume that $i=b_1b_2$. Then $i c_{00}=b_1b_2 \,1=c_{11}$ whose quantum counterpart should read $i|00\rangle=|11\rangle$, making
$|00\rangle$ and $|11\rangle$ linearly dependent. We have to proceed differently.

A way out of the difficulty was proposed in \cite{C07} and is based on $i$ defined by (\ref{i}). Intuitively, this $i$ is equivalent to a $\pi/2$ rotation in a real plane (the convention used in \cite{C07} differed by a sign from (\ref{i}), but we prefer the latter choice).
The additional bit $A_0$ introduces the doubling of the dimension analogous to the one associated with real and imaginary parts of a single complex number.

Our definition also implies that $i^2=-1$ so that
\be
e^{i\phi}c_{0_0A_1\dots A_n}
&=&
\cos\phi\, c_{0_0A_1\dots A_n} + \sin\phi\, i c_{0_0A_1\dots A_n}\\
&=&
\cos\phi\, c_{0_0A_1\dots A_n} + \sin\phi\, c_{1_0A_1\dots A_n}
\ee
has all the required properties of, simultaneously, a complex number multiplied by a complex phase factor, and a 2-dimensional real vector rotated by an angle $\phi$.

Let us illustrate these considerations by a GA representation of the state
\be
|\psi\rangle = \frac{1}{\sqrt{2}}
\big(
|10\rangle
+
e^{i\phi}
|01\rangle\big).
\ee
Now we have to use three bits and a three dimensional space spanned by $b_0$, $b_1$, and $b_2$.
The GA analogue reads
\be
\psi
&=&
\frac{1}{\sqrt{2}}
\big(
c_{010}
+
e^{i\phi}
c_{001}\big),\\
&=&
\frac{1}{\sqrt{2}}
\big(
c_{010}
+
\cos\phi\, c_{001}+ \sin\phi\, ic_{001}\big)\\
&=&
\frac{1}{\sqrt{2}}
\big(
c_{010}
+
\cos\phi\, c_{001}+ \sin\phi\, c_{101}\big)\\
&=&
\frac{1}{\sqrt{2}}
\big(
b_1
+
\cos\phi\, b_2+ \sin\phi\, b_{02}\big)\label{20}.
\ee
Obviously, the multivector (\ref{20}) can be easily visualized in various ways.

\subsection{Scalar product}

Scalar product does not seem important for GA computation (we do not really need `bras'). But just for the sake of completeness let us mention the following construction.

Consider a comb $c_{A_1\dots A_n}=b_1^{A_1}\dots b_n^{A_n}$. Its {\it reverse\/} is
$c_{A_1\dots A_n}^*=b_n^{A_n} \dots b_1^{A_1}$. The geometric product
$
c_{A_1\dots A_n}^*c_{B_1\dots B_n}
$
equals 1 if and only if $(A_1,\dots,A_n)=(B_1,\dots,B_n)$. If the two sequences of bits are not identical, the product
$
c_{A_1\dots A_n}^*c_{B_1\dots B_n}
$
is a blade different from 1. Let now $\Pi_0$ denote the projection of a multivector on the scalar part
$1=c_{0\dots 0}$. Then
\be
\Pi_0\,c_{A_1\dots A_n}^*c_{B_1\dots B_n}=\delta_{A_1B_1}\dots \delta_{A_nB_n}.
\ee
The latter formula might be used to define a GA scalar product, if needed.

\subsection{Elementary gates}

GA analogues of elementary gates (Pauli, Hadamard, phase, $\pi/8$, cnot, and Toffoli) were described in \cite{C07}. Here we want to shed some light on the issue of how many elementary operations are associated with networks of gates.

We have to begin with yet another additional dimension, represented by the basis vector $b_{n+1}$. This additional dimension will not be used for coding, but for defining certain bivector operations. We do it as follows \cite{C07}. Let $a_j=b_jb_{n+1}$, $0\leq j\leq n$, and consider any complex --- in the sense of (\ref{complex c}) --- vector $z_{A_1\dots A_k\dots A_n}$.
Negation of a $k$th bit, $1\leq k\leq n$,
\be
{\rm n}_kz_{A_1\dots A_k\dots A_n}
&=&
b_k\Big(\prod_{j=0}^{k-1}a_j\Big)^*z_{A_1\dots A_k\dots A_n}\prod_{j=0}^{k-1}a_j\nonumber\\
&=&
z_{A_1\dots A'_k\dots A_n},
\ee
is defined in purely algebraic terms. The same with another important operation, multiplication by $(-1)^{A_k}$,
\be
a_k^* z_{A_1\dots A_k\dots A_n}a_k
&=&
(-1)^{A_k}z_{A_1\dots A_k\dots A_n}.
\ee
All the elementary one-, two-, and three-bit gates can be defined in terms of
${\rm n}_k$, $(-1)^{A_k}$, and $i$ \cite{C07}. Of particular importance is the linear map
\be
{\rm A}_k z_{A_1\dots A_k\dots A_n}
&=&
\frac{1}{2}
z_{A_1\dots A_k\dots A_n}
-
\frac{1}{2}
a_k^* z_{A_1\dots A_k\dots A_n}a_k\nonumber\\
&=&
A_kz_{A_1\dots A_k\dots A_n}.
\ee
Now let $X$ be any map of GA into itself satisfying $X(0)=0$. Then $X_k=1-{\rm A}_k+X\circ {\rm A}_k$ is a control-$X$, controlled by the $k$th bit. Let us note that ${\rm A}_k$ is a projector on the subspace spanned by those blades that contain the vector $b_k$. In particular, in Fig.~5 the selection of blades containing the red $\nearrow$ is performed, algebraically speaking, by means of ${\rm A}_8$.

In order to understand how to count the number of algorithmic steps let us take a concrete example of, say, $n$ Hadamard gates acting on different bits.
Let $\psi$ be a multivector,  not necessarily a single blade, but a general combination of $2^n$ possible blades. The Hadamard gate simultaneously affecting the $k$th bits of all the $2^n$ blades of $\psi$ is \cite{C07}
\be
H_k \psi
&=&
\frac{1}{\sqrt{2}}
\big(
{\rm n}_k \psi
+
a_k^* \psi a_k
\big).
\ee
Similarly to $\psi$, $H_k\psi$ is a {\it single\/} multivector.

The latter observation is trivial, perhaps, but crucial for the problem.
A simple illustration of a 2-bit multivector $\psi=\psi_0 1+ \psi_1 b_1+ \psi_2 b_2+ \psi_{12} b_1b_2$ is a 2-dimensional oriented plane segment (represented by $\psi_{12} b_1b_2$), suspended at the height $\psi_0$, and whose center of gravity is above the point $\psi_1 b_1+\psi_2 b_2$. A gate maps $\psi$ into some new $\psi'$ which has a similar geometric interpretation.

So when it comes to the question of how many operations are performed while computing $H_1\dots H_n\psi$, the answer is this: Two for computing
$H_n\psi$, another two for computing $H_{n-1}H_n\psi$, yet another two for computing $H_{n-2}H_{n-1}H_n\psi$, and so on. Finally, we need $2n$ operations. The same argument applies to all the other elementary gates.

If we do not know how to treat multivectors as single geometric objects, then computing $H_1\dots H_n\psi$ will involve an exponential number of steps. So the key ingredient of efficient geometric-algebra computation is to implement all the needed gates in a geometric manner.
In Section IV~C we give a concrete example of realization of $H_1\dots H_n\psi$ in $2n$ steps, if $\psi=c_{0_1\dots 0_n}$.

\subsection{Reading superposed information}

An important difference between quantum and geometric coding is in the ways one gets information from superpositions of states. In quantum coding a measurement projects a superposition on a randomly selected basis vector. Such measurements destroy the original state. In GA coding the ontological status of superpositions is different. Here one has a collection of geometric objects and thus can perform many measurements on the same system without destroying its state. As a consequence certain standard ingredients of quantum algorithms are not needed in GA-based computations.

Shor's algorithm \cite{Shor} for factoring 15 into $3\cdot 5$ provides a simple illustration. The entangled state
$|\psi\rangle=\sum_{x=0}^{15}|x\rangle |a^x\,{\rm mod}\,15\rangle$, for $a=2$, reads
\be
|\psi\rangle&=&
\big(|0\rangle+|4\rangle+|8\rangle+|12\rangle\big) |1\rangle\nonumber\\
&+&
\big(|1\rangle+|5\rangle+|9\rangle+|13\rangle\big) |2\rangle\nonumber\\
&+&
\big(|2\rangle+|6\rangle+|10\rangle+|14\rangle\big) |4\rangle\nonumber\\
&+&
\big(|3\rangle+|7\rangle+|11\rangle+|15\rangle\big) |8\rangle\nonumber.
\ee
The goal is to find the period of the function $x\mapsto 2^x\,{\rm mod\,}15$, but the problem is that the sequences of numbers correlated with the values $2^0=1$, $2^1=2$, $2^2=4$, $2^3=8$, are periodic, but shifted by, respectively, 0, 1, 2, and 3. To get rid of this shift one performs quantum Fourier transformation on the first register.

We claim that in a GA version of the algorithm the Fourier transform step will not be needed. What we have to do is to localize the vectors
$|x\rangle|1\rangle$ and the smallest $x>0$ is the solution. The GA analog of the calculation is given by the multivector
\be
{}&{}&
\big(c_{0_10_20_30_4}+c_{0_11_20_30_4}+c_{1_10_20_30_4}+c_{1_11_20_30_4}\big) c_{0_50_60_71_8}\nonumber\\
&{}&\pp++
\nonumber\\
&{}&
\big(c_{0_10_20_31_4}+c_{0_11_20_31_4}+c_{1_10_20_31_4}+c_{1_11_20_31_4}\big) c_{0_50_61_70_8}\nonumber\\
&{}&\pp++
\nonumber\\
&{}&
\big(c_{0_10_21_30_4}+c_{0_11_21_30_4}+c_{1_10_21_30_4}+c_{1_11_21_30_4}\big) c_{0_51_60_70_8}\nonumber\\
&{}&\pp++
\nonumber\\
&{}&
\big(c_{0_10_21_31_4}+c_{0_11_21_31_4}+c_{1_10_21_31_4}+c_{1_11_21_31_4}\big) c_{1_50_60_70_8}\nonumber.
\ee
The cartoon version of this computation is shown in Fig.~5. Selecting an appropriate subset of shapes we find that the period is $x=4$. The factorization is given by $2^{x/2}\pm 1$.
We have not needed the Fourier transform. An analogous observation was made in the context of the Simon algorithm in \cite{MO}.

\begin{figure}
\includegraphics[width=5 cm]{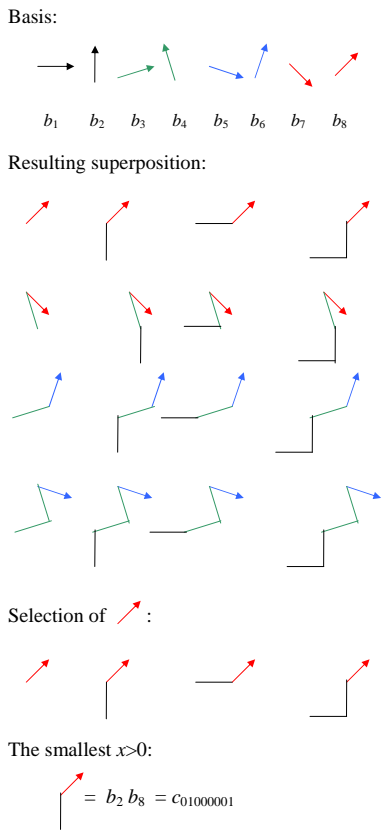}
\caption{(Color online) Shor-type algorithm. Euclidean space is 8-dimensional. Oracle first produces the multivector playing a role of the entangled state. Then a selection of blades containing the red $\nearrow$ is performed. The blade $c_{01000001}$ corresponding to the smallest number $x>0$ is selected. The first half of bits represents $x=4$.}
\end{figure}

\subsection{Mixed states}

The example of the Shor algorithm shows clearly that
multivector coefficients, as opposed to wave functions, do not have a probabilistic interpretation. For this reason the GA analogues of quantum algorithms are not probabilistic. Still, probabilistic algorithms will occur if one replaces multivectors by multivector-valued random variables.
The resulting states will be mixed in the usual meaning of this term (probability measures defined on the set of pure states) but nevertheless will not, in general, lead to a density matrix formalism (the latter occurs only in theories where pure-state averages of random variables are bilinear functionals of pure states).

In this context we should mention the paper \cite{Christian} where multivector-valued hidden variables were used to violate an analogue of the Bell inequality. The construction employs a hidden-variable state that is mixed in our sense. Although the problem posed in \cite{Christian} is not exactly equivalent to the one addressed in the original Bell construction \cite{Bell}, it is nevertheless interesting from our point of view, and shows a way of generating certain GA analogues of quantum correlation functions.

\subsection{GA versions of quantum algorithms}

We will not give here explicit GA versions of quantum algorithms, since separate papers \cite{AC07,MO,Pawlowski} were devoted to this subject.
The general conclusion is that any quantum algorithm has a GA analogue, a fact following from the GA construction of the elementary quantum gates \cite{C07}. The main formal difference between GA and quantum computation is that the GA formalism is not bound to use unitary gates. Indeed, the unitarity of quantum gates follows from the Schr\"odinger equation formula $U_t=\exp (-iHt)$, which is irrelevant for GA computation. What {\it is\/} relevant in the GA framework are those operations that have a geometric meaning. In particular, one of the most important non-unitary geometric operations is a projection on a subspace. This projection has nothing to do with the projection postulate of quantum mechanics. Indeed, in quantum mechanics the projection `collapses' a superposition on a basis vector. The geometric projection projects a multivector on another multivector, but in general not on a single comb.

A situation where geometric projections play a simplifying role in GA analogues of quantum algorithms is the problem of deleting intermediate `carry bits' in quantum adder networks \cite{E,Preskill,VMeter,Cheng}. A glimpse at the network proposed in \cite{E} shows that the number of gates could be reduced by almost a half if one did not insist on performing this task in a reversible way. This is especially clear if one compares alternative adder
networks discussed in \cite{Cheng}.

In terms of GA computation this concrete part of the network will be replaced by an appropriate projection which, in spite of being irreversible, is geometrically allowed. For example, in a 3-bit case the operation of resetting the third bit, $c_{ABC}\to c_{AB0}$, corresponds to the following set of projections
\be
1=c_{000} &\to& c_{000}=1,\\
b_1=c_{100} &\to& c_{100}=b_1,\\
b_2=c_{010} &\to& c_{010}=b_2,\\
b_3=c_{001} &\to& c_{000}=1,\\
b_{1}b_{2}=c_{110} &\to& c_{110}=b_{1}b_{2},\\
b_{1}b_{3}=c_{101} &\to& c_{100}=b_{1},\\
b_{2}b_{3}=c_{011} &\to& c_{010}=b_{2},\\
b_1b_{2}b_{3}=c_{111} &\to& c_{110}=b_1b_{2}.
\ee
Each of them has a geometric interpretation: $b_1b_{2}b_{3} \to b_1b_{2}$ squeezes a cube into its $x-y$ wall; $b_{2}b_{3} \to b_{2}$ squeezes a square lying in the $y-z$ plane into its side parallel to the $y$ axis, and so on. An interpretation in terms of the polylines is left to the readers.

\section{Multivector analogues of important pure states}

Let us finally give explicit multivector counterparts of some important entangled states occurring in
quantum information problems.

\subsection{Bell basis}

The Bell basis consists of four mutually orthogonal 2-qubit entangled states:
\be
|\Psi_\pm\rangle
&=&
\frac{1}{\sqrt{2}}\big(|01\rangle
\pm
|10\rangle\big),\\
|\Phi_\pm\rangle
&=&
\frac{1}{\sqrt{2}}\big(|00\rangle
\pm
|11\rangle\big).
\ee
There are two bits and thus a 2-dimensional Euclidean space will suffice as long as we do not need complex numbers.
Let the basis be $b_1$, $b_2$.
The corresponding multivectors then read
\be
\Psi_\pm
&=&
\frac{1}{\sqrt{2}}\big(c_{01}
\pm
c_{10}\big)=\frac{1}{\sqrt{2}}\big(b_2
\pm
b_1\big),\\
\Phi_\pm
&=&
\frac{1}{\sqrt{2}}\big(c_{00}
\pm
c_{11}\big)
=
\frac{1}{\sqrt{2}}\big(1
\pm
b_{12}\big).
\ee
Fig.~6a shows the corresponding sets of blades. The blades $\Psi_\pm$ are represented simply by two unit vectors rotated by $\pm\pi/4$ with respect to the axis spanned by $b_1$. It is interesting that an analogous simple representation of an entangled state occurs in quantum optics formulated in the so-called $\infty$-representation of canonical commutation relations \cite{PC}. The two basic blades correspond there to two modes of light behind a beam splitter.
\begin{figure}
\includegraphics[width=8 cm]{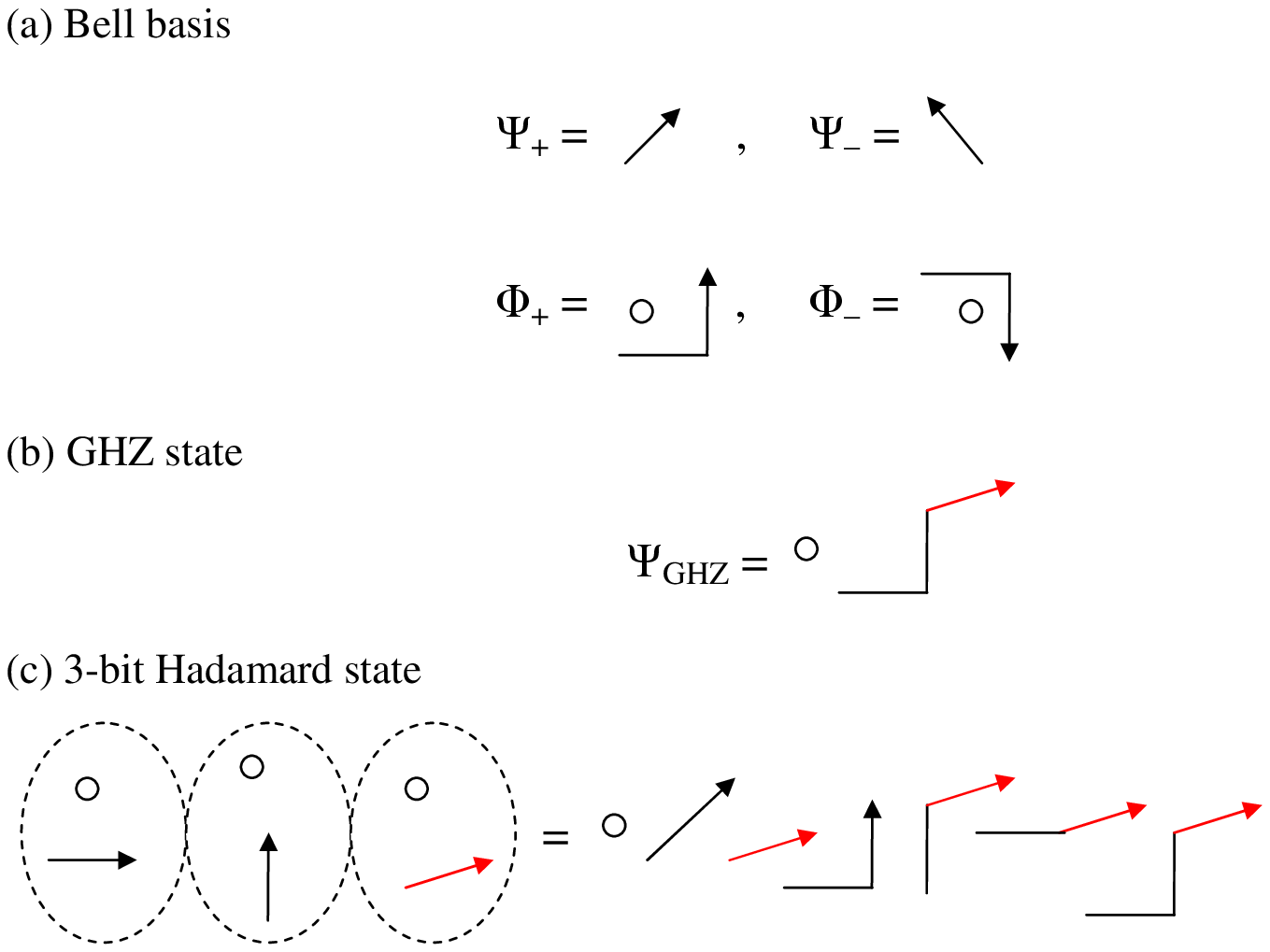}
\caption{(Color online) Cartoon versions of entangled states: (a) The Bell basis, and (b) the 3-bit GHZ state. (c) Action of a 3-bit Hadamard gate on $c_{000}$. The combination $b_1+b_2$ is shown as a single black vector. All the blades are assumed to be appropriately normalized.}
\end{figure}
\subsection{GHZ state}

The 3-bit GHZ state reads
\be
|\Psi_{\rm GHZ}\rangle=
\frac{1}{\sqrt{2}}\big(
|000\rangle +|111\rangle\big).
\ee
The GA representation
\be
\Psi_{\rm GHZ}=
\frac{1}{\sqrt{2}}\big(
c_{000} +c_{111}\big)
=
\frac{1}{\sqrt{2}}\big(
1 +b_{123}\big)
\ee
is shown in a cartoon form in Fig.~6b.

\subsection{$n$-fold Hadamard state}

An $n$-fold Hadamard state is obtained if one acts with an $n$th tensor power of a Hadamard gate on a `vacuum' $|0\dots 0\rangle$. As a result one gets a superposition of all the $n$-bit numbers $|A_1\dots A_n\rangle$. Such a state is the usual starting point for quantum computation. In the GA formalism the corresponding multivector reads \cite{C07}
\be
2^{-n/2}(1+b_1)\dots (1+b_n)=2^{-n/2}\sum_{A_1\dots A_n}c_{A_1\dots A_n}.
\ee
Fig.~6c shows its 3-bit illustration.

Let us note that the superposition of $2^n$ combs $c_{A_1\dots A_n}$, representing all the $n$-bit numbers, is here obtained by means of $n$ aditions and $n$ multiplications. This step is as efficient as its quantum version and agrees with our previous analysis of the $n$-fold Hadamard gate in Section III~D.

\section{Conclusions}

It seems fair to say that quantum computation looks from the GA perspective as a particular implementation of a more general way of computing. The implementation based on tensor products of qubits and quantum superposition principle is characteristic of the quantum world. However, the formalism of quantum computation loses its micro-world flavor when viewed from the GA standpoint. Actually, there is no reason to believe that quantum computation has to be associated with systems described by quantum mechanics. GA occurs whenever some geometry comes into play. It is enough to thumb the monographs of the subject \cite{H1,HS,H2,Baylis,S,DDL,Pavsic,Doran} to understand its ubiquity, interdisciplinary character, and vast scope of applications.

The question of concrete practical implementation of GA coding is an open one and is certainly worth of further studies.

\acknowledgments
This work was supported by the Flemish Fund for Scientific Research (FWO), project G.0452.04. MC thanks J. Rembieli\'nski for support and encouragement,  and Z. Oziewicz for comments.

\end{document}